\documentclass[manuscript]{acmart}
\usepackage[option]{anonymous-acm}
\usepackage{subcaption}
\begin{document}

\title{Dancing Through Soundscapes: Designing a Low-Cost, Sound-Based Device for Sensing and Interpreting Movement and Dance}

\author{Swen E. Gaudl}
\affiliation{%
\institution{University of Gothenburg}
\city{Gothenburg}
\country{Sweden}
}
\additionalaffiliation{%
\institution{Seam CoLab Ltd.}
\city{Bristol}
\country{United Kingdom}
}
\email{swen.gaudl@gu.se}
\orcid{0000-0003-3116-3761}

\author{Silvia Carderelli-Gronau}
\affiliation{%
\institution{BathSPA University}
\city{Bath}
\country{United Kingdom}
}
\additionalaffiliation{%
\institution{Seam CoLab Ltd.}
\city{Bristol}
\country{United Kingdom}
}
\email{m.carderelli-gronau@bathspa.ac.uk}

\renewcommand{\shortauthors}{Gaudl \& Carderelli-Gronau}

\begin{abstract}
When we move through space, we often rely on multiple senses beyond vision to perceive and act in that environment: we ``feel'' the presence of others; we build internal representations and models and recall them to navigate the environment. We also leave traces and impressions that others pick up on. The traces include echoes, heat, the displacement of objects such as furniture or footprints, air movement close to the face of another, smells such as perfume, but also the immediate sounds we make when we move and breathe. Movement is a spatial and temporal activity, and dance as a form of movement practice requires coordination of oneself in relation to others, the space and a potential score. When rehearsing dance, dancers have to relate to others often not just by looking but more often by feeling and imagining or remembering where others are based on experience and shared practice. So how can we approach technology-mediated movement and dance? Why should we explore it? How can we support spatial and temporal practice meaningfully and joyfully? In this work, we focus on sound traces; we present the design and rationale for a sound-based artefact that translates movement-based sound into layered, explorable, generative soundscapes. The work contributes a novel artefact for exploring movement-based activities with a audio-first approach, with a focus on the spatial performative experience. The paper further reflects on observations from workshops and public sharings, including how participants used repetition, stillness, environmental sound, and call-and-response to understand and improvise with the soundscape.

\end{abstract}

\begin{teaserfigure}
\centering
\includegraphics[width=\textwidth]{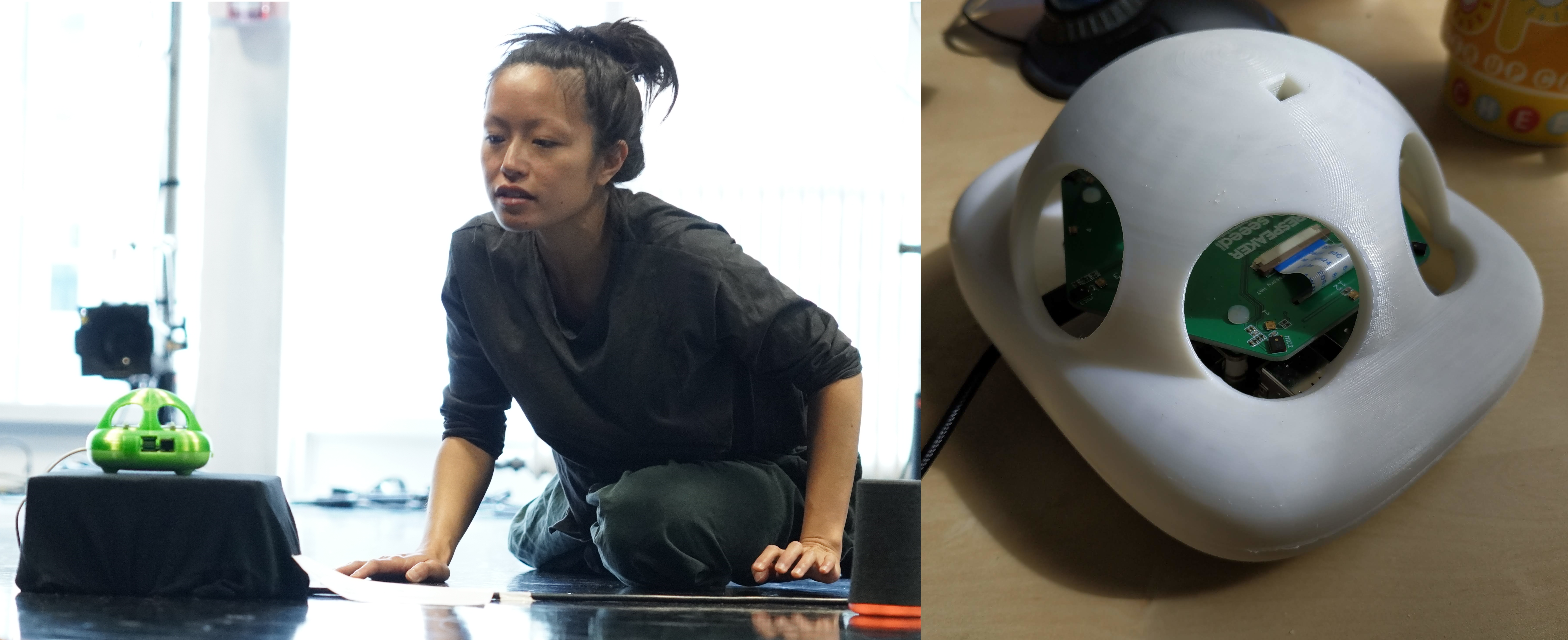}
\caption{Left: Dancer discussing their experience with the device (green artefact); Right: Close-up of the device.}
\Description{Left image: A female dancer in the centre-right of the image is leaning forward, actively discussing the artefact with a person in the lower-left. Right image: A white device with large round holes that show the electronics inside.}
\label{fig:teaser}
\end{teaserfigure}


\maketitle

\section{Introduction}

Activities or domains that depend on embodied presence, shared physical space, or nuanced interpersonal proximity still provide open challenges that should be explored \cite{zhou_dance_2021}, with opportunities for translating these practices into online formats that preserve their core qualities. Dance is one such form of activity. Dance is expressive communication in which spatial and temporal movement qualities convey information, relations, intentions, feelings, and affect using its own language. It is not solely directed towards a single other entity but addresses different others. It addresses other dancers but also the audience.

Due to its strong spatial and embodied nature, the audience experiences the sound, rhythm, kinaesthesia, architecture, and affect of the performance. However, conventional mediated approaches focus on visual representation. For dancers who are part of the performance, the perceived experience extends beyond the visual, as they feel the presence of others based on where people are, where they should be based on practice and knowledge about the other, and the traces they leave. Thus, this is an experience that is not traditionally captured.

Movement-based practices rely on spatial awareness, dynamic relationality, and a constant negotiation of distance, rhythm, and presence between movers. Most screen-based telepresence systems \cite{hove2022shortcomings} offer only a limited and flattened representation of these relationships. Movers must orient themselves around a fixed visual point, the screen, interpreting others' actions through a single point of view while trying to infer spatial relationships that are normally felt through embodied cues \cite{dourish2001action}. Although some systems attempt to map participants into shared virtual planes, the underlying interaction remains visually constrained, reducing dancers' ability to perceive where others are in relation to their own bodies.

The opportunities and limitations of video-based tools during the COVID-19 pandemic made these challenges increasingly apparent \cite{bench2021we}. Across artistic and educational settings, dancers, teachers, and facilitators reported fatigue, challenges in maintaining shared spatial awareness, interpreting partners' movement intentions, and sustaining a sense of embodied connection when relying solely on screens \cite{serdar2024clash}.

Even after the pandemic, remote collaboration in creative endeavours such as dance practice, rehearsal, or performance is worth facilitating where travel or co-location is not possible. The presented work integrates the development of a novel technical artefact into the existing domain of mediated and augmented performative art. It focuses on sound as a medium through sonification, spatial audio, and generative soundscapes, not to complement a performance, but to support dancers in engaging with one another and to provide an additional approach for communicating presence, proximity, and relational movement.

The design vision was to imagine a system that would allow movers to \emph{feel} each other across distance without depending on vision \cite{kiridoshi2022spatial}.

In this paper, we present a description of an interactive artefact, as well as a reflection on the design process and observations of its use. Through these reflections, we identify the artefact's responsiveness, soundscape quality, physical robustness, and sensitivity to environmental sound as central to how participants formed individual and shared movement relationships with it.


\section{Related Work}

\subsection{Movement, Dance, and Embodied Interaction}

In the context of dance and performance, sonification can shape dancers' bodily awareness and artistic practice through gesture--sound interaction \cite{dahlstedt2019otokin,giomi2020somatic}. These works explore a co-evolution of choreographic practice and interactive technology, highlighting the aesthetics of sensation and perception. Interactive dance technologies and new musical interfaces have explored how dancers can shape sound through wearables, sensing surfaces, and bespoke instruments, often emphasising performance-led research and first-person perspectives.

During the pandemic, a strong push towards mediated education also reached into the domains of arts and dance, changing the way dance could be taught and explored \cite{li2020teaching}. Some of the benefits of this new creative space remained, but the move towards employing videoconferencing technology was driven mostly by necessity rather than through co-creating and speculating about new possibilities.

A final body of work we link to is in the domain of soma, somatics \cite{hanna1986somatics,green2002somatic}, and somaesthetic research \cite{hook2018designing}, where the focus is on sensing, organising, and knowing movement from within. Embodiment asks how dance is lived, relational, cognitive, cultural, and political. Thus, when thinking about how a dancer acts and interacts, we do not only need to consider the external traces they create, such as movement-based sounds, but also what the experience feels like and how their interaction within a space and with their body shapes their own internal state and perception.

\subsection{Movement Sonification and Generative Sound}

Movement sonification research investigates how human movement data can be transformed into sound to support perception, motor control, learning, and rehabilitation \cite{effenberg2016movement,sonification_motor_learning}. By providing sensory feedback for a movement action or a rhythmic pattern of actions, studies show that additional sonification can improve the coordination of movement and motor learning beyond rhythmic cueing alone, highlighting the potential of sound as an informative feedback channel \cite{sonification_motor_learning}. This literature offers conceptual and technical tools for mapping movement to sound that informed some of our generative mapping strategies.

Generative soundscapes and ambient generative audio have been explored in sound art, acoustic ecology, and interactive media for decades, including early work on real-time generative soundscape creation and more recent overviews of generative sound design that connect rule-based and machine-learning approaches \cite{birchfield2005design,Menexopoulos2023state}. Visual programming environments such as Max and Pure Data have long supported custom generative audio systems, which provide an important technical platform for exploring sound generation as a design material \cite{puckette2007theory,jekosch2005assigning}.

In the domain of computational music, interactive music systems for dance and embodied performance are a recurring theme. The \emph{Sensitiv} project, for example, presents a sonic co-play tool designed from the perspective of a dancer and musician, supporting interactive dance through sound feedback \cite{anderssonlopez2021_sensitiv}. Other research explores generative soundscapes and embodied control, such as installations where breath control shapes generative forest soundscapes or where textile-based interfaces support interactive dance instruments \cite{gifford2025_breathing_with_the_forest,madaghiele2024_pain_creature,nabi2024_embodied_exploration_latent_spaces}.

\subsection{Spatial Audio and Dance Technology}

Within the HCI field, there is a growing body of work on generative and spatialised audio, as well as dance and movement technologies. Spatial audio has been used to enhance immersive visual environments and data exploration; for example, \emph{Audible Panorama} automatically generates spatial audio for 360-degree panorama images to increase the realism of VR scenes \cite{huang2019_audible_panorama}. More recent work explores spatial audio for multimodal graph rendering and learning on touchscreen devices, as well as its effects on social presence in a space \cite{robinsonmoore2024_spatial_audio_graph_rendering,kiridoshi2022spatial}.

Researchers have proposed spatialised audio interactions with segmented audio in immersive environments for engaging with performing-arts intangible heritage, highlighting how spatial audio can support active engagement and situated experience \cite{wang2025_from_temporal_to_spatial}. This work resonates with the artefact described here due to its emphasis on bodily engagement with cultural and artistic practices, while differing in its use of sound-based sensing rather than primarily visual or pre-authored content.

Dance- and movement-focused research has examined dance instruction in VR, anticipatory movement visualisation, and technology-mediated audience interaction in dance performance, identifying challenges and best practices for designing for dancers, audiences, and creative processes \cite{laattala2024_wave,correia2024_best_practices_audience_interaction}. These works tend to focus on visual instruction and audience-facing interaction. The presented approach complements them by offering an audio-first, screen-reduced approach to shared dance.

Our work draws inspiration from these examples but departs from movement-tracking approaches that rely on cameras or body-worn sensors by using sound-based localisation and foregrounding low-cost and replicable hardware.


\section{The Artefact}

\subsection{Design Rationale}

The design and technical development of \textanon{\textit{SonicDancer}}{the artefact} evolved through research through design \cite{zimmerman2007research}, co-creation, and speculative design and prototyping \cite{dunne2024speculative}. The detailed co-design process and workshop chronology are outside the scope of this paper, as this process spanned four years and multiple iterations, only selected design-process reflections can be covered within the presented work.

The initial design exploration focused on how remote movement practices might be supported without restricting the scope to current technological products. Key questions included:

\begin{itemize}
\item How can you dance together when you cannot be in the same physical space?
\item How can you touch somebody to feel or detect their presence?
\item What qualities can exist in a virtual space?
\item How can you coordinate spatial movement?
\end{itemize}

The team looked into qualities that the physical space provides and envisioned remote touch, where a dancer could feel the touch of someone or where they could reach into space to feel the presence of another dancer. For contemporary dance, sensing proximity and reaching out to touch somebody were key qualities that emerged through the design exploration.

When envisioning ensemble movement, the idea of a connected network of people moving together through a shared experience emerged, but went beyond the existing shared screen in current videoconferencing solutions. Instead, the design imagined a shared virtual space where movers could experience the relation to others through and within their bodies.

Virtual Reality headsets were considered impractical because they are strapped to the body, affecting possibilities for free movement, such as fast head movement or moving on the floor. VR headsets were also seen as precious by dancers. When moving, people considered how they could damage the technology, foregrounding the needs of the technology.

Additional sensors, including motion tracking, touch-sensitive devices, and force-feedback vests used in gaming, also emerged from the initial discussions. These were seen as less accessible for students and non-academics and have already been partially explored for music composition \cite{kanga2024reflections}. Videoconferencing also creates a window into people's homes, which works for some forms of practice but can be invasive.

The idea of a technology that connects people across spaces, is not worn on the body, is less invasive, and provides some form of touch or spatial connection through different sensory modalities emerged from the design phase.

For this work, we use the term \emph{virtual touch} to describe two bodies occupying the same virtual space and receiving sound feedback. The aim is not to reproduce tactile contact directly, but to explore a sound-based indication of presence and spatial relation.

\subsection{Physical Artefact}

\begin{figure}
\centering
\begin{subfigure}{0.6\textwidth}
\includegraphics[width=\textwidth]{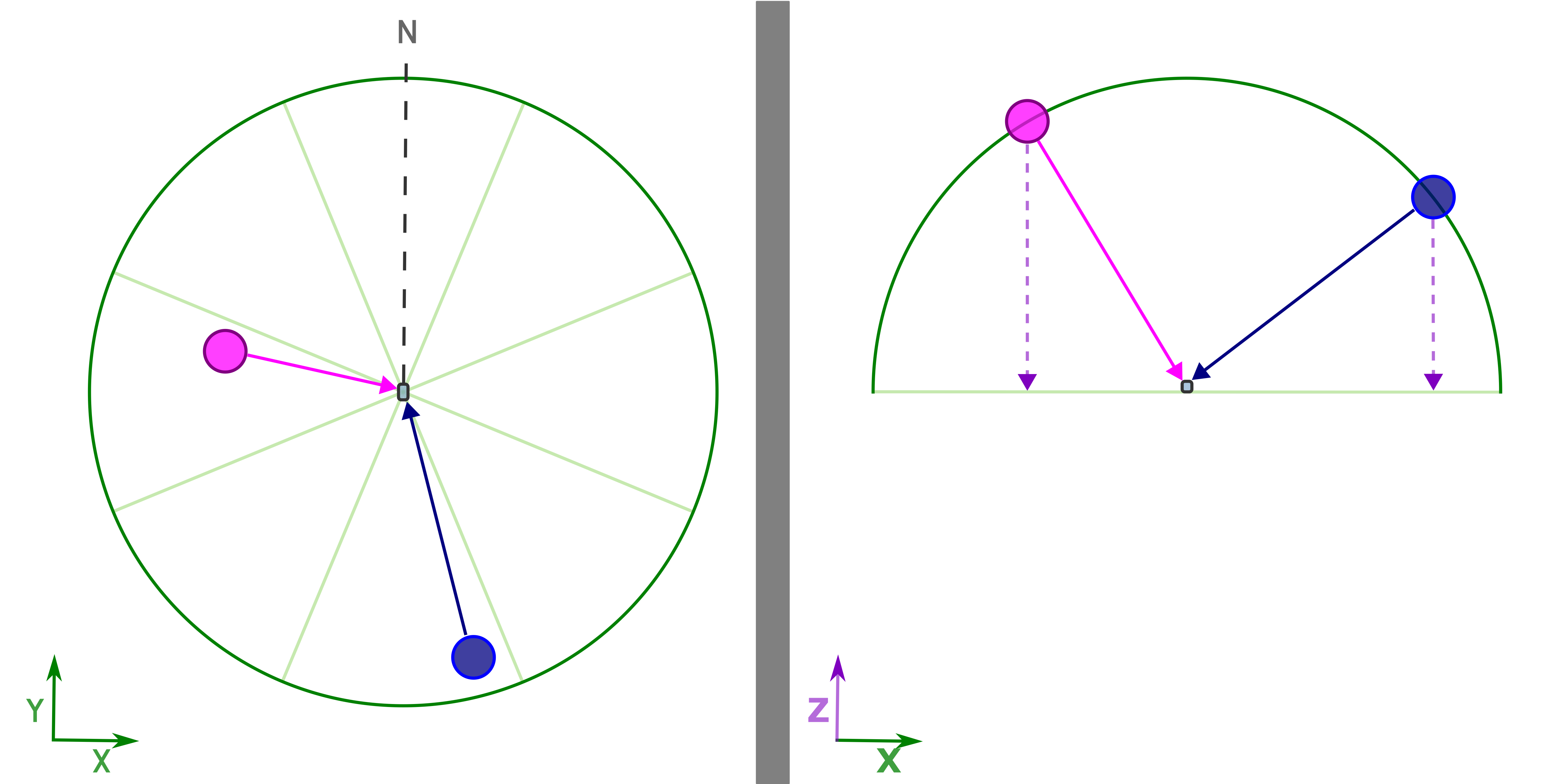}
\caption{Spatial layout with the XY-plane on the left and the XZ-plane on the right. The green circle represents the performance space with a radius of $r=1m$. The device is at the centre, and two movers are in the space.}
\label{fig:spatial}
\end{subfigure}
\hfill
\begin{subfigure}{0.35\textwidth}
\includegraphics[width=\textwidth]{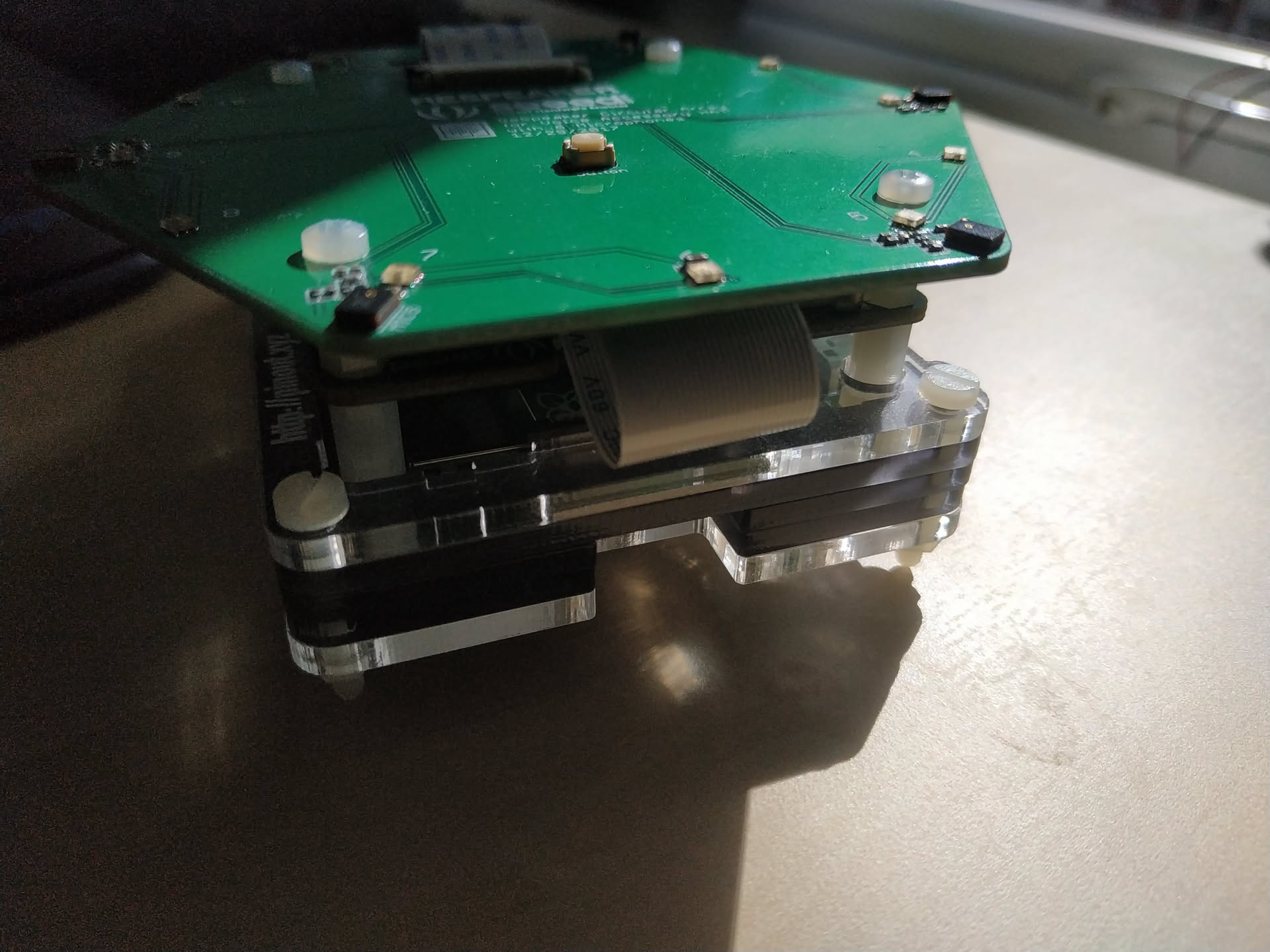}
\caption{The Raspberry Pi 4, including the microphone array, outside the shell.}
\label{fig:SD-open}
\end{subfigure}
\caption{The device's spatial layout, as well as a closer look at the device without its shell.}
\label{fig:artefact}
\end{figure}

\textanon{\textit{SonicDancer}}{The artefact} is positioned within a movement space. It is contained inside a robust 3D-printed shell with rounded corners, as shown in Figure~\ref{fig:teaser}, and creates minimal spatial obstruction when positioned within the space. The enclosure protects the electronics, as shown in Figure~\ref{fig:SD-open}, and the mover. It is mechanically robust enough to be stepped on, bumped into, or kicked during dance.

The device weighs in at $weight_{w/}=349g$ with the shell and $weight_{w/o}=97g$ without the shell. Its dimensions are $width_{w/}=130mm$, $depth_{w/}=150mm$, $height_{w/}=100mm$, and $width_{w/o}=101mm$, $depth_{w/o}=115mm$, $height_{w/o}=35mm$. The shell is printed from PETG, a weather-resistant and sturdy plastic. We currently have multiple printed shells in different colours to allow users to customise their devices and to make it easy to identify a specific device. Some variations are printed from semi-transparent PETG to increase the effect of the LEDs, while others are printed from a more matt material that has a smoother touch.

The device is intended to provide a fixed position, like a lighthouse, at the centre of the movement space; see Figure~\ref{fig:spatial}. Movers can move around it or over it while the microphone array detects sound produced within the surrounding area. The minimum space, based on our workshops with dancers, is ideally a circular environment that has a radius of $r_{min}=1m$, the green circle in Figure~\ref{fig:spatial}; another example is visible in Figure~\ref{fig:dancing}, where the tape markers form a circle. However, we also explored larger spaces of up to $r_{max}=5m$, which work well. As the microphone picks up directions in 3D space\footnote{The 3D directions are mapped on a sphere and for simplification projected onto the ground plane giving us direction and elevation during processing.}, a circle is the ideal shape; however, the performance space can be any shape when working with the devices. The circle is internally split into segments. In Figure~\ref{fig:spatial}, you can also see two tracked entities moving in the space. On the left side you see a top-down view with the directions that the device picks and the underlying segments a sound source is assigned to. In this example two tracked entities are shown. On the right side, elevation is shown as the dotted lines going down from an the entities.

In terms of connections or wires going through the space, the device only requires a USB-C power supply, which is covered and secured with tape to the floor. This reduces the risk of tripping but also makes it easier to return the device to its position if it is kicked. A device can also be powered by a phone power bank for hours. In the current configuration, the device is used with either Bluetooth speakers or Bluetooth headphones. The speakers are initially positioned next to the device, but the movers are free to use them as an extension of their body. In some settings and workshops we encouraged them to pick up and use them actively in others dancers automatically picked up the speakers. The speakers provide not only auditive feedback but, due to some of the audio output we designed being bass-heavy, also haptic feedback.

\subsection{Interaction Design Goals}

Exploring the interaction design of \textanon{\textit{SonicDancer}}{the artefact} narrowed the questions to focus on space, embodiment, sensory modalities, and experience. This led to four key design goals:

\begin{itemize}
    \item \textbf{Embodied, not screen-centric:} Participants should attend primarily to their bodies, the surrounding space, and the connection they form with the soundscape, rather than to a screen.
    \item \textbf{Low entry threshold, high expressivity:} Interaction should be accessible without training, yet reward continued exploration with nuanced control.
    \item \textbf{Shared presence:} The soundscape should make other movers perceptible as situated, dynamic presences, locally and remotely.
    \item \textbf{Inclusivity:} Interaction should be possible without reliance on vision and should be open to different physical abilities.
\end{itemize}

\subsection{Interaction Possibilities}

In this section, the two most common modes of using the device are explained: local interaction and remote interaction (pair). More modes are also possible, including connecting more than two devices, using visual representations of the data, or broadcasting the output before or after the soundscape is generated.

\begin{figure}[h]
\centering
\includegraphics[width=0.5\linewidth]{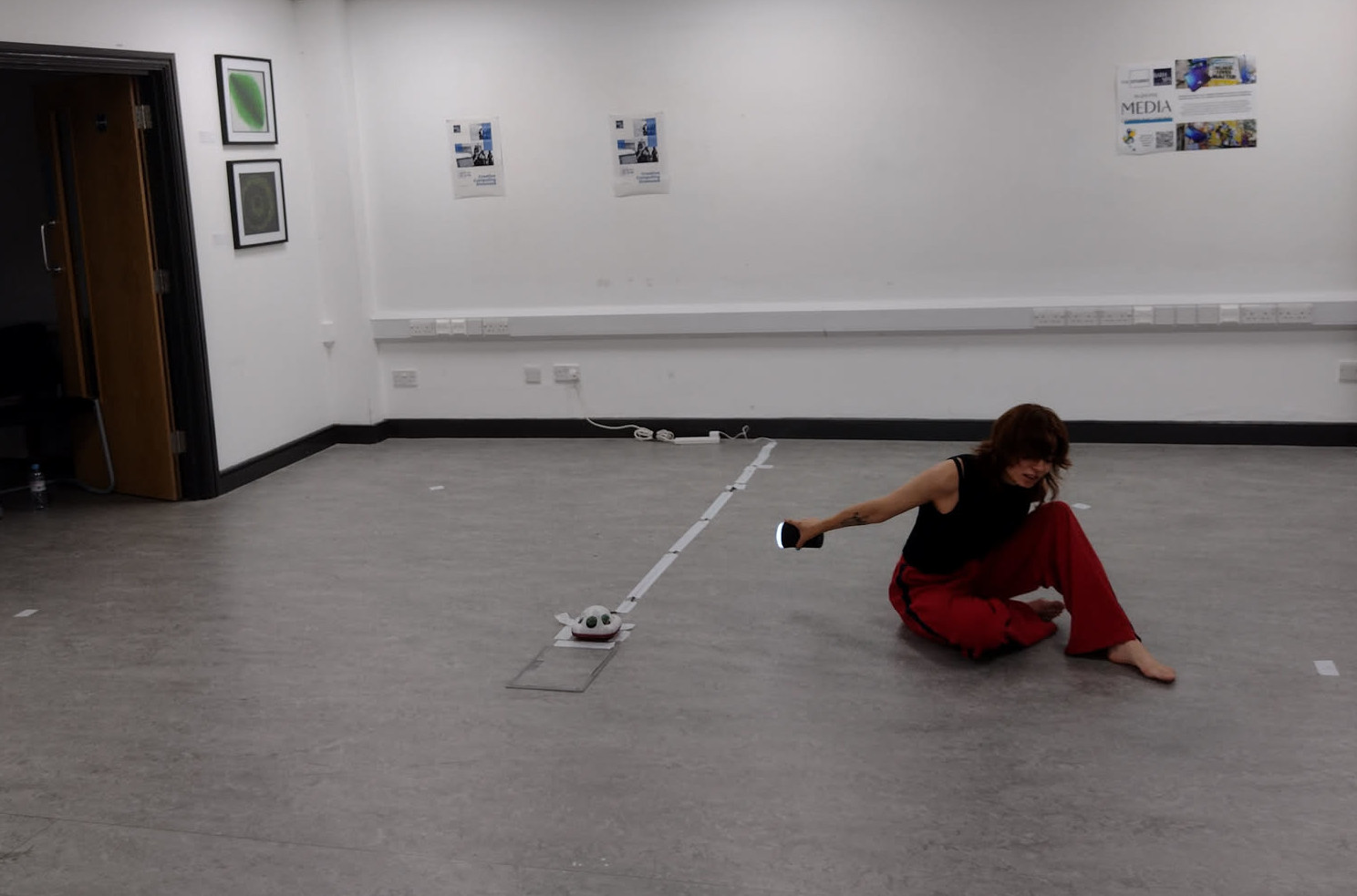}
\caption{A dancer exploring local interaction with the device, using a Bluetooth speaker as output and tracking the ambient sound of the room, including the generated positional spatialised soundscape.}
\label{fig:dancing}
\end{figure}

\paragraph{Local Interaction:}

In a local configuration, one or more participants move within the space around the device. Each device uses the direction it faces as a baseline. The space should fit a circle with a radius of $r_{min}=1m$. Their footsteps, weight shifts, and other movement-generated sounds are captured and localised. The resulting trajectories are mapped into an internal representation, which in turn drives the generative sound engine.

Participants experience changes in the soundscape as they move around the device: their ``sonic voice'' may shift in timbre or register when crossing sector boundaries from segment to segment on the ground plane; increased movement energy may thicken the sound or introduce additional layers; stillness may thin or quiet the soundscape. Since the mappings are continuous rather than discretely triggered, small variations in movement produce perceptible but subtle sonic differences, encouraging sustained improvisation.

The LED ring provides optional orientation cues, particularly during initial familiarisation or for participants with partial sight, but is not necessary for ongoing interaction. In Figure~\ref{fig:dancing}, a dancer is using the local configuration to explore the soundscape and her responses. She is using a Bluetooth speaker as a part of her body, not simply to listen to the sound in a different relation to her body, but also to provide an additional source, extending her capability to trigger the device.

\begin{figure}
\centering
\includegraphics[width=0.95\linewidth]{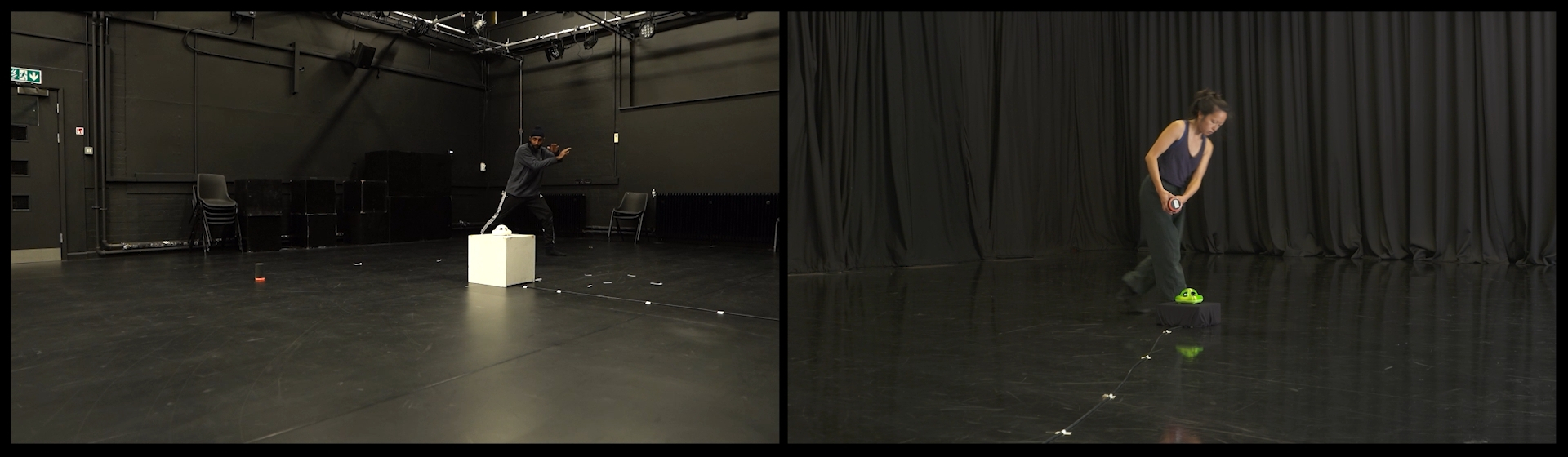}
\caption{Two dancers are improvising together in a remote setting. Each dancer is listening to the sound that the other dancer generates after having first explored how the device works and responds in local mode.}
\label{fig:paired-movement}
\end{figure}

\paragraph{Remote Interaction:}

In a remote configuration, at least two devices are installed in different locations, each with local participants; see Figure~\ref{fig:paired-movement}. Each device shares its movement representation over Wi-Fi and renders a soundscape either combining local and remote trajectories or only representing remote trajectories. Participants in separate spaces thus hear each other as part of the same generative environment without sharing video. In this setting, Bluetooth speakers can be used, but headphones create a clearer relation between the movement of a dancer and their soundscape.

This arrangement supports remote ensemble improvisation and invites participants to explore how their movement can respond to and influence the soundscape produced by others. Due to the nature of the data, it is possible to assign dedicated qualities to individual spaces, making the combination more transparent, like an orchestra of different instruments. Currently, assigning dedicated qualities or instruments has not yet been explored.

\subsection{Spatial Representation and Soundscape}

The system listens to sound generated within the space around the device and renders it as a dynamic soundscape. Ambient sound captured by the microphone array is processed using ODAS \cite{grondin2022odas}, an open-source direction-of-arrival and sound-source-tracking library. Rather than identifying bodies or gestures, the system estimates where generated sounds originate in relation to the device. As part of this, ODAS is used as a first stage, returning a continuous list of all identified sound sources, as well as a list of tracked identified sources. These are then internally stored, filtered, and clustered every few hundred milliseconds ($i=375ms$).

The result of ODAS is then mapped onto the segments of the ground plane, each representing a clearly differentiable sound quality. In the current version, the segments are mapped to a harmonic scale, turning each segment into a piano-like key. Through our workshops, we set up an orientation on the device, similar to a compass that points north, so the mover has an understanding of where each key is; see the segments in the left sketch in Figure~\ref{fig:spatial}. This direction (north) can be adjusted within the program at runtime. The idea of the piano-like key is not to simply trigger a MIDI note, but also to use, as inputs in addition to the segment, the certainty of the algorithm regarding where a user moved, the elevation of the source above the ground, and the strength of the source to create a sound response within the soundscape, similar to what can be done with MIDI Polyphonic Expression (MPE)\footnote{\url{https://midi.org/mpe-midi-polyphonic-expression}}, which, in terms of piano playing, creates a more natural and expressive style of playing instead of just turning sound on or off. To reduce the impact of noise or unwanted ambient sounds, the device also allows the user to set a threshold below which sounds are ignored. This allows the device to be used in both noisy and quiet environments, depending on the choice of the user. The result for the soundscape is up to 10 spatialised sources that provide a rich representation to the generator. As the sound changes with the user moving, the sound is spatialised in that respect. Distance from the device is currently not measured or inferred, as the loudness of a sound does not directly relate to the distance from the device, considering that the sound is not always produced with the same intensity.

The source representation is translated into a generative soundscape implemented in Pure Data. Instead of triggering only discrete sonic events, the sound engine modulates textures and densities and layers sounds based on the detected sources.

\subsection{Hardware Architecture}

The current \textanon{\textit{SonicDancer}}{device} integrates the following main components:

\vspace*{2ex}
\textbf{Single-board computer:} A Raspberry Pi, currently a Raspberry Pi~4 with 4GB of RAM, serves as the primary processing unit, running the entire stack (localisation, processing, distribution, and audio generation) on Patchbox OS\footnote{\url{https://blokas.io/patchbox-os}}. Price: $\approx$ 100 euros.

\textbf{Microphone array:} A circular multi-microphone array, the six-microphone ReSpeaker board\footnote{\url{https://wiki.seeedstudio.com/ReSpeaker_6-Mic_Circular_Array_kit_for_Raspberry_Pi/}}, is mounted on the device, providing multichannel audio input suitable for direction-of-arrival estimation, as well as the LED array detailed below. It is no longer available, but newer arrays such as the ReSpeaker 4 work as well. Price: $\approx$ 50 euros.

\textbf{Bluetooth USB Dongle/Audio:} The device connects to Bluetooth speakers or headphones. This choice supports deployments ranging from studio use to small home environments. Additionally, keeping it wireless works well for dancers. Dancers can also use their own headsets, so there are no extra costs. An additional Bluetooth 5.0 USB dongle is used with the Raspberry Pi, as the internal Bluetooth occasionally generates issues and interferes with the Wi-Fi. Using the dongle also improved pairing with headphones and speakers due to compatibility issues with the original Bluetooth chip on the Pi. Price: $\approx$ 10 euros.

\textbf{Enclosure:} A 3D-printed shell protects the electronics and is mechanically robust enough to be stepped on and bumped during dance. The shells are designed for user customisation, so users can switch the top and bottom parts to create colour variations. The design was created through testing with multiple dancers, including a visually impaired dancer who preferred bright options that could be illuminated by the LEDs. Price: $\approx$ 5 euros.

\textbf{LED compass:} The LED array is shaped in a ring and is used to provide a low-resolution, peripheral visual cue indicating the relative directions of other movers without requiring a separate screen. It colour-codes tracked movers, so if multiple movers are in the space and move, they are represented by a fixed colour.

\textbf{Power:} The device runs from a USB-C power supply or battery pack, supporting portable and outdoor use. When using the device outside, or if a power socket is not available, a normal phone power bank provides enough electricity to power the device for over an hour.

This hardware configuration balances robustness, cost, and accessibility, making it possible for artists and community organisations to assemble and deploy their own devices. The low-cost nature of having a device at $\approx$160 euros and the robust design also make it less precious, so people can freely explore the possibilities without being afraid of breaking it.

\subsection{Software Architecture}

To integrate the hardware components and manage the device, the software is split into different layers with separate functionality. An introduction and setup configuration guide is available in our project repository \textanon{\textit{[anon]}}{\footnote{\url{github.com/GU-IxD-AI/SonicDancer/}}}. The software is freely available and contains a stable version.

The software stack is organised into four layers:

\begin{enumerate}
\item \textbf{Sound capture, localisation, and processing:} Multichannel audio from the microphone array is processed using ODAS, an open-source sound-source-localisation library \cite{grondin2022odas}. In the current configuration, ODAS estimates the direction of arrival of a sound (DOA). It generates a list of sources and a list of tracked sources from prior tracking and streams these continuously into a processing component running on the same device. During processing, the original localised sources are analysed and clustered. If a source is below the listening threshold, it is removed before it feeds into the clustering. The processing stage also keeps its own log of previously tracked clusters and blends the ODAS-tracked sources only partially to increase tracking accuracy. The result of the processing is up to 10 tracked sources representing entities that produce sound in the environment.

\item \textbf{Movement representation.} A middleware layer aggregates localisation data and maps it into the onion-like representation composed of angular sectors with the additional qualities of elevation, signal strength, tracking ID, and tracking certainty. This stage also assigns unique colours to the \emph{tracked entities}\footnote{\emph{Tracked entities} are sound sources that have persisted over a number of cycles of processing and relate to a more consistent entity in the space that can be stationary of mobile.}, which are shown on the LED compass. Tracking, conversion of data, and retrieving the processed data from the lower layer are completed in under half a second; this was derived through experimentation to balance tracking quality and responsiveness of the soundscape. This layer maintains high-level objects corresponding to detected sources and handles smoothing.

\item \textbf{Generative sonification engine.} The movement representation is streamed into Pure Data patches that implement the generative sound engine. Each sector is associated with a spectral, textural, or rhythmic character. Movement speed, persistence, and dispersion modulate parameters such as density, filtering, and pitch regions. The generator also controls the layering and blending of sector-based audio.

\item \textbf{Networking and remote coupling.} Multiple \textanon{\textit{SonicDancer}}{devices} may connect over Wi-Fi using sockets (TCP/IP) and share their high-level movement representations. No audio data are shared, but a compressed representation of the sound sources and the tracked entities is shared. Thus, each device could perform additional processing on on a lower level, closer to the ambient audio, of a connected device or only use the already processed tracked entities. Each device then integrates local and remote movement into a configurable generative soundscape that blends all connected devices.
\end{enumerate}

The system prioritises processing on the Raspberry Pi and transmits movement descriptors rather than full audio between devices. All processing and computation, as well as the generation of audio, are done on the device, creating a fully encapsulated and compact solution.

\section{Reflections on the Artefact}

Over the duration of four years, the presented artefact has been continuously iterated upon, from the initial idea and design speculation to a working technical device that can be manufactured in maker spaces. The project was started as a collaboration between a computer scientist and a dancer/movement therapist, with the vision of exploring technical approaches that are beneficial for supporting the movement practice of multiple dancers while dealing with the restrictions of the COVID-19 pandemic. After initial sessions between the initiators of the project, the project was scaled up through workshops involving professional dancers, students (dance, music and computing), and the general public to explore the design space and develop something that provided value to them. A total of ten workshops were carried out over the first three years, alongside additional sharing sessions with audiences. 

Within those ten workshops, a total of 37 individuals (6 dancers, 2 sound designers, 2 computer scientists, and 27 students) participated; on average we had seven participants per workshop. The workshops were either one or two day events. We also shared the artefact with the public in four larger events with over 350 attendees. For all sessions we collected informed consent to record and share the results and the collected material. 

A key focus of the project was not to let technology get in the way of the practice, the experience, and a person's moving body by adding technology just for its own sake or out of curiosity. The primary focus was on the dancers, their interaction, and their experience.

The project started with a design speculation that led to a first physical prototype through remote collaboration. Under COVID-19 restrictions, the authors met in person for the first time after half a year, and the dancer explored the recorded spatial audio and its translation into a 3D visual representation. The resulting discussions centred on what the actual soundscape of a room looks like, similar in some aspects to Deep Listening \cite{oliveros2005deep}.

The team then explored possible ways of making this information intentionally explorable, as derived from the design sessions de-focusing screen-based solutions, leading to the idea of sonification and having an audio designer react to the visual representation, following a Wizard of Oz approach. After a number of sessions with the audio designer imitating what automation would do, a first automated version was derived based on feedback from the audio designer, the dancers' responses, and iterative prototyping tests.

In the following reflections, changes and takeaways from the process of making and breaking the artefact are presented as observations of the process rather than as results or formal evaluations.

\subsection{Encountering the Device}

During the first sessions with people using the device, we used the Wizard of Oz approach, and dancers explored the space with a visual representation on a projector, while the audio designer used the same representation to generate live soundscapes. After the sessions, the resulting discussions with the dancers centred on the concepts of presence and on how the visual representation drew attention towards the screen, while it also took them some time to form a concept of the audio and how it related to movement. It was also difficult for the audio designer to express their rationale or processes for creating audio live in response to the visuals, so we focused on co-creating the soundscape by iterating the technical implementation and audio design while testing with dancers. All further sessions were carried out with a fully generative soundscape.

When the screen was turned off, dancers explored the space more freely, not paying particular attention to anything outside the space when placing and interacting with the speaker. Dancers, including a visually impaired dancer, also commented on the shell being nearly invisible in the space during movement and felt comfortable that it should not be treated as something precious. In a couple of cases, dancers kicked and stepped on it, which did not destroy the device or hurt them, reassuring them.

When dancers explore the device for the first time, they create loud sounds by vocalising, clapping, and making heavy footsteps. After a while, they realise that the device also picks up the sound of breathing, light footsteps, the sound of their clothing, and touching the floor, and their approach to movement shifts. Some of them intentionally mix louder and softer sounds or vocalisation, while others explore the limits of the space and the device in terms of fast and slow movement or distance from it. After becoming accustomed to the device and space, dancers mentioned that, when performing a score or exploring the soundscape, the device faded into the background and they could concentrate either on their own sound or on connecting to the paired space.

One senior dancer commented that they were initially disappointed by the lack of accuracy when moving through the space and that gestures and postures were not captured. After exploring the soundscape, the response dramatically shifted, which motivates our continued work on the artefact. What they commented was that the device captures a form of movement quality beyond posture tracking, focusing on the body moving through space and preserving some of the intentionality. They also felt that the soundscape as a medium was beneficial to capturing dance, as dancers respond to music.

When non-dancers explored the device, they also started with clapping, finger snaps, and heavy footwork. Most of them then started exploring the soundscape by circling around the device. When offered a way of exploring the device without anyone watching, some people responded that they felt they could connect to themselves and came out of the space relieved. All users so far have mentioned that they felt the device was either connecting to them or, when using remote interaction, that somebody was there and interacting with them.

\subsection{Movement and Sound}

While the audio designer was creating sound live, it was harder to explore repeated movement and sound responses, which had been the team's intention from the start. Once the generative soundscape was working, we could also see how people perceived the connection between movement, sound, and response. Most of the dancers, when forming their model of the soundscape, explored repeated movements, such as visiting a spot in the space multiple times or exploring different poses. Initially, some of the dancers were disappointed that poses did not register, but they then started exploring the 3D aspect of the space, as well as the different qualities of movement, more.

Over the course of the project, we also changed the soundscape multiple times, from a few simple oscillators towards complex, textured, and themed patterns, and designed sound palettes that can be altered. With the simpler palette, dancers could quickly understand the space but found the soundscape too harsh and annoying after a while, a response we did not receive with the new textured palettes. In a few cases, we also had intense responses, such as crying with joy, due to the call-and-response nature of local interaction, where the device responds to the mover.

We also had dancers explore in more depth the difference between movement/stillness and sound/silence, which was rarely done by non-dancers. Interestingly, the sensitivity and the ambient soundscape played an important role in understanding the space. If the environmental responses from the device were initially too strong, dancers could not form a mental model of the space, as the device seemed ``blind'' to them. However, once they were used to the device, bringing in more environmental sounds was interesting to some dancers, as it provided a background track of their individual space.

When bringing in more dancers, we also saw that some actively used the environment to produce sounds in addition to their bodies or the speaker. Some dancers vocalised while moving, which works as an additional strong signal. However, dancers used this intentionally, so talking or singing while moving did not create an issue.

The final observation was that both the response time and the quality and ``feeling'' evoked by the soundscape had the greatest impact on the movement carried out by dancers and non-dancers.

\subsection{Moving with Others}

When in remote interaction mode, dancers explored different ways of interacting. Dancers initially listened to the sound, trying to identify where people were in the space, which only worked well with lower environmental disturbances. One approach was call-and-response: one dancer would move, and the other tried to respond and move in the same way. An observed extension of this was dancers trying to occupy the same spot, creating what we envisioned as virtual touch. Throughout all sessions, this was explored by dancers either with or without prompting.

When using simple oscillators, the touch resulted in a unison of sound waves. Other patterns included doing the opposite and trying to avoid the other dancer. After becoming more accustomed to the soundscapes, dancers started exploring moving together, not by following these fixed patterns, but by exploring and improvising together using the soundscapes as a shared canvas.

We only explored non-dancers listening to dancers in remote interaction after they had explored the soundscapes themselves. They commented that they enjoyed listening and that they tried to understand what the dancers were doing and where they were in the space.

\subsection{Limitations and Breakdowns}

The artefact is not intended as a replacement for existing technology, as it cannot solve all challenges in remote movement practices. The artefact does not replace posture or gesture tracking but focuses on capturing spatial movement with an intentionally simple setup.

As the device is ``blind'' if the environmental sounds are too loud, it will not pick up movers, so noise has a large impact. However, with visual tracking, the same applies in very bright or very dark spaces. Our current setup was tested outside with natural background noise, as well as in a fully dark environment, and it captured dancers well after the sensitivity was adjusted. Environmental sounds have been appreciated by dancers in our workshops after they developed an understanding of how the device works. That said, when unanticipated disturbances occur during a session, such as construction work at one site, they can lead to confusion in the other remote space.

To address technical challenges such as network loss or headphones powering down, we implemented a simple start-up procedure in the device so that, upon rebooting, it resumes its prior state and restarts everything. Thus, when a device crashes during a session, simply ``turning the device off and on again'' works well.

One of the current challenges is setting up the spaces for dancers and offering an easy way to connect devices. The instructions require some technical knowledge, but the team is working towards a simplified app that allows setup, adjustment, and ways of connecting to and storing experiences.

\section{Contributions of the Artefact}

\textanon{\textit{SonicDancer}}{The artefact} contributes to tangible and embodied interaction and the interactive arts along several dimensions:

\begin{itemize}

\item It demonstrates the use of sound-based localisation for embodied interaction without cameras or body-worn sensors, offering an untethered exploration of the space the the soundscape unfolds.

\item It offers a generative soundscape approach that encodes movement and spatial relations rather than primarily speech, posture, gesture, or symbolic control.

\item It provides a screen-reduced interaction modality through which dancers and non-dancers can explore movement-generated sound individually and across remotely connected spaces.

\item It follows a maker-style approach to the rapid development of physical-digital artefacts, combining a Raspberry Pi, microphone array, 3D-printed enclosure, ODAS-based localisation, and Pure Data sound generation in a single low-cost and physically robust device.

\item It documents how participants formed an understanding of the soundscape through repeated movement, contrasts between movement and stillness, environmental sound, and call-and-response.

\item It contributes reflections on the importance of engaging with movers (dancers and non-dancers) and investigates response time, soundscape quality, sensitivity to environmental sound, and physical robustness when designing sound-based artefacts that are intended to foster movement practice rather than draw attention away from it.

\item It documents how dancers explored remote interaction through call-and-response, avoidance, shared improvisation, and attempts to occupy the same represented position, described in this work as virtual touch.
\end{itemize}

\section{Conclusion and Future Work}

We have presented \textanon{\textit{SonicDancer}}{the artefact} as a low-cost, sound-based device for shared movement through generative live soundscapes. The system combines a small single-board computer with maker-scene components and a generative audio system to render movement-generated sound as an evolving soundscape.

The artefact explores an interaction that reduces reliance on screens, cameras, and body-worn sensors. It uses sounds produced through movement as input for representing spatial activity and generating sound. Across workshops and public sharings, dancers and non-dancers initially explored the system through loud, easily identifiable sounds before exploring quieter bodily and environmental sounds, moving from obvious to subtle. Participants used repetition, movement/stillness, call-and-response, avoidance, and shared improvisation to understand and interact through the soundscape.

The reflections indicate that the physical robustness of the device and the removal of a required screen allowed the artefact to recede into the background as participants became familiar with it. At the same time, more technical and design aspects such as response time, soundscape quality, sensitivity to environmental noise, and the ease of setting up and connecting devices substantially shaped the interaction.

Future work will improve the setup and connection process, provide simpler controls for sensitivity and environmental sound, and extend the networking architecture to support larger ensembles. We plan to explore the generative audio space as well as the different use cases for archiving performances in more detail. Considering the feedback from the mixed-ability users, we also plan to deepen the collaboration with visually impaired and mixed-ability dancers and develop travelling installation formats, such as small movement-and-listening stations for community venues.

\bibliographystyle{ACM-Reference-Format}
\bibliography{sonicdancer,temp}

\end{document}